\pdfoutput=1
\RequirePackage{ifpdf}
\ifpdf 
\documentclass[pdftex]{sigma}
\else
\documentclass{sigma}
\fi

\DeclareMathOperator{\tr}{tr}

\makeatletter
\newcommand{\fmslash}[2][0mu]{%
  \mathchoice
    {\fmsl@sh\displaystyle{#1}{#2}}%
    {\fmsl@sh\textstyle{#1}{#2}}%
    {\fmsl@sh\scriptstyle{#1}{#2}}%
    {\fmsl@sh\scriptscriptstyle{#1}{#2}}}
\newcommand{\fmsl@sh}[3]{%
 \m@th\ooalign{$\hfil#1\mkern#2/\hfil$\crcr$#1#3$}}
\makeatother
\numberwithin{equation}{section}

\begin{document}

\newcommand{\arXivNumber}{1402.6184}

\allowdisplaybreaks

\renewcommand{\thefootnote}{$\star$}

\renewcommand{\PaperNumber}{054}

\FirstPageHeading

\ShortArticleName{Two-Point Functions on Deformed Spacetime}

\ArticleName{Two-Point Functions on Deformed Spacetime\footnote{This paper is a~contribution to the Special Issue on
Deformations of Space-Time and its Symmetries.
The full collection is available at \href{http://www.emis.de/journals/SIGMA/space-time.html}
{http://www.emis.de/journals/SIGMA/space-time.html}}}

\Author{Josip TRAMPETI\'C~$^{\dag\ddag}$ and Jiangyang YOU~$^\dag$}

\AuthorNameForHeading{J.~Trampeti\'c and J.~You}

\Address{$^\dag$~Rudjer Bo\v skovi\' c Institute, P.O.~Box 180, HR-10002 Zagreb, Croatia}
\EmailD{\href{mailto:josipt@rex.irb.hr}{josipt@rex.irb.hr}, \href{mailto:youjiangyang@gmail.com}{youjiangyang@gmail.com}}

\Address{$^\ddag$~Max-Planck-Institut f\"ur Physik (Werner-Heisenberg-Institut),\\
\hphantom{$^\ddag$}~F\"ohringer Ring 6, D-80805 M\"unchen, Germany}

\ArticleDates{Received February 24, 2014, in f\/inal form May 19, 2014; Published online May 29, 2014}

\Abstract{We present a~review of the one-loop photon ($\Pi$) and neutrino ($\Sigma$) two-point functions in a~covariant and
deformed $\rm U(1)$ gauge-theory on the 4-dimensional noncommutative spaces, determined by a~constant antisymmetric tensor
$\theta^{\mu\nu}$, and by a~parameter-space $(\kappa_f,\kappa_g)$, respectively.
For the general fermion-photon $S_f(\kappa_f)$ and photon self-interaction $S_g(\kappa_g)$ the closed form results reveal
two-point functions with all kind of pathological terms: the UV divergence, the quadratic UV/IR mixing terms as well as
a~logarithmic IR divergent term of the type $\ln(\mu^2(\theta p)^2)$.
In addition, the photon-loop produces new tensor structures satisfying transversality condition by themselves.
We show that the photon two-point function in the 4-dimensional Euclidean space\-time can be reduced to two f\/inite terms by
imposing a~specif\/ic full rank of $\theta^{\mu\nu}$ and setting deformation parameters $(\kappa_f,\kappa_g)=(0,3)$.
In this case the neutrino two-point function vanishes.
Thus for a~specif\/ic point $(0,3)$ in the parameter-space $(\kappa_f,\kappa_g)$, a~covariant $\theta$-exact approach is able to
produce a~divergence-free result for the one-loop quantum corrections, having also both well-def\/ined commutative limit and
point-like limit of an extended object.}


\Classification{81R60; 81T10; 81T15}

\renewcommand{\thefootnote}{\arabic{footnote}}
\setcounter{footnote}{0}

\section{Introduction}

Prior to the late 1990s, the possibility of experimentally testing the nature of quantum gra\-vi\-ty was not seriously contemplated
because of the immensity of the Planck scale ($E=1.2
\times
10^{19}$~GeV).
Now this view has been signif\/icantly alternated: A possibility to have a~string scale signif\/icantly below the Planck scale in
a~braneworld scenario~\cite{ArkaniHamed:1998rs} become the core of current experimental protocols searching for quantum-gravity
phenomena (notably production of black holes) at the Large Hadron Collider at CERN.
Almost simultaneously, another possibility arose, where it was pointed out that distant astrophysical objects with rapid time
variations could provide the most sensitive opportunities to probe {\it very high energy} scales, i.e., almost the near-Planck
scale physics~\cite{AmelinoCamelia:1997gz}.

Another route to search for quantum-gravity ef\/fects involves a~{\it spontaneous breaking of Lorentz symmetry} in string theory,
when a~tensor f\/ield acquires a~vacuum expectation value (vev).
Unlike the case of scalars, these tensor vevs do carry spacetime indices, causing the interaction represented by the Standard
Model (SM) f\/ields coupled to these vevs to depend on the direction or velocity of the said f\/ields.
Stated dif\/ferently, these background vevs bring about the breakdown of Lorentz symmetry.
This entails a~distinctive fact of most of Lorentz violating (LV) theories on the existence of {\it preferred reference frames},
where the equations of motion take on the simplest form.
In contrast to the notion of the {\it motionless aether} from the end of the 19th century, we have a~rather unique example of
such a~frame in modern cosmology today: the frame in which the Cosmic Microwave Background Radiation (CMBR) looks isotropic.
From the determination of the detailed spectrum of the CMBR dipole (generally interpreted as a~Doppler shift due to the Earth's
motion), our velocity with respect to that frame, of order of $10^{-3}$~c, can be inferred.

An eligible way to infer the preferred reference frame predicted by generic quantum gravity frameworks, is to study dispersion
relations for propagating particles.
Instead of propagating (in a~vacuum) with the speed of light, in Lorentz violating theories one expects an energy-dependent
velocity~${\bf v}(E)$ for massless particles.
This is a~consequence of the loss of Lorentz covariance in the dispersion relations for propagating particles, with the
implication that a~specif\/ic form~${\bf v}(E)$ can be at best valid only in one specif\/ic reference frame.
Thus, a~preferred reference frame in which the equation of motions possess the simplest form is singled out.
This opens up a~unique possibility to study constraints on violations of Lorentz invariance.
The modif\/ication of the photon velocity of the form~${\bf v}(E)$ would induce time lag for photons of dif\/ferent energies, which
could be subsequently detected if such particles can propagate at cosmological distances.
Such an alternation of the photon velocities has already been obtained in {\it Loop quantum gravity} (being another popular
approach to quantum gravity)~\cite{Rovelli:2010bf,Rovelli:2011eq} as well as in heuristic models of space-time foam inspired by
string theory~\cite{AmelinoCamelia:1997gz}.

One of the most striking observation regarding spontaneous Lorentz breaking via tensor vevs in the string theory framework is
that it can be formulated as deformed f\/ield theories.
Specif\/ically, a~low-energy limit is identif\/ied where the entire boson-string dynamics in a~Neveu-Schwartz condensate is
described by a~minimally coupled supersymmetric gauge theory on noncommutative (NC) space~\cite{Seiberg:1999vs} such that the
mathematical framework of noncommutative geometry/f\/ield theory~\cite{Connes,Douglas:2001ba,Madore,Szabo:2001kg} does apply.
In such a~scenario, noncommutative Dirac--Born--Infeld (DBI) action is realized as a~special limit of open strings in a~background~$B^{\mu \nu}$ f\/ield, in which closed string (i.e.\
gravitational) modes are decoupled, leaving only open string interactions.
Since in string theory~$B^{\mu \nu}$ f\/ield is a~rather mild background, the antisymmetric tensor $\theta^{\mu \nu}$ governing
spacetime noncommutative deformations is not specif\/ied, and therefore the scale of noncommutativity could, in principle, lie
anywhere between the weak and the Planck scale~\cite{Bigatti:1999iz, Douglas:2001ba,Szabo:2001kg,Szabo:2009tn}.
It is thus of crucial importance to set a~bound on this scale from experiments~\cite{Szabo:2009tn}.

It is important to stress the invariance of the theory under coordinate changes, i.e., the invariance under an observer
transformation (where the coordinates of the observer are boosted or rotated).
This transformation is not related to the concept of Lorentz violation since in this transformation the properties of the
background f\/ields transform to a~new set of coordinates as well.
On the other hand, an invariance under ${\it active}$ or ${\it particle}$ transformation, where both f\/ields and states are being
transformed, is now broken by the background f\/ields themselves, leading to the concept of Lorentz violation.

In a~simple model of the noncommutative spacetime we consider coordinates $x^\mu$ as the Hermitian operators $\hat
x^\mu$~\cite{Jackiw:2001jb},
\begin{gather}
[\hat x^\mu,\hat x^\nu]= i\theta^{\mu\nu},
\qquad
|\theta^{\mu\nu}|\sim\Lambda^{-2}_{\rm NC},
\label{4dim}
\end{gather}
where $\theta^{\mu\nu}$ is {\em a~constant} real antisymmetric matrix of dimension $length^2$, and $\Lambda_{\rm NC}$ being the
scale of noncomutativity.
It is straightforward to formulate f\/ield theories on such noncommutative spaces as a~deformation of the ordinary f\/ield
theories~\cite{Bigatti:1999iz, Douglas:2001ba,Szabo:2001kg,Szabo:2009tn}.
The noncommutative deformation is implemented by replacing the usual pointwise product of a~pair of f\/ields~$\phi(x)$ and
$\psi(x)$ by the star($\star$)-product in any action:
\begin{gather*}
\phi(x)\psi(x) \longrightarrow (\phi\star \psi)(x)=\phi(x)\psi(x) + {\cal O}(\theta,\partial \phi,\partial \psi).
\end{gather*}
The specif\/ic Moyal--Weyl $\star$-product is relevant for the case of a~constant antisymmetric noncommutative deformation tensor
$\theta^{\mu\nu}$ and is def\/ined as follows:
\begin{gather}
(\phi\star \psi)(x)= e^{\frac{i}{2}\theta^{\mu\nu}{{\partial}^\eta_\mu} {{\partial}^\xi_\nu}}
\phi(x+\eta)\psi(y+\xi)\big|_{\eta,\xi\to0} \equiv\phi(x) e^{\frac{i}{2}\overleftarrow{{\partial}_\mu}
\theta^{\mu\nu}\overrightarrow{{\partial}_\nu}} \psi(x).
\label{f*g}
\end{gather}
The $\star$-product has also an alternative integral formulation, making its non-local character more transparent.
The coordinate-operator commutation relation~\eqref{4dim} is then realized by the star($\star$)-com\-mu\-ta\-tor of the usual coordinates
\begin{gather*}
[\hat x^\mu,\hat x^\nu]=[x^\mu \stackrel{\star}{,} x^\nu]=i\theta^{\mu\nu},
\end{gather*}
implying the following spacetime uncertainty relations
\begin{gather*}
\Delta x^\mu \Delta x^\nu \geq \frac{1}{2}|\theta^{\mu\nu}|.
\end{gather*}
The above procedure introduces in general the f\/ield operators ordering ambiguities and breaks ordinary gauge invariance.

Since commutative local gauge transformations for the D-brane ef\/fective action do not commute with $\star$-products, it is
important to note that the introduction of $\star$-products induces f\/ield operator ordering ambiguities and also breaks ordinary
gauge invariance in the naive sense.
However both the commutative gauge symmetry and the deformed noncommutative gauge symmetry describe the same physical system,
therefore they are expected to be equivalent.
This disagreement is remedied by a~set of nonlocal and highly nonlinear parameter redef\/initions called Seiberg--Witten (SW)
map~\cite{Seiberg:1999vs}.
This map promotes not only the noncommutative f\/ields and composite operators of the commutative f\/ields, but also the
noncommutative gauge transformations as the composite operators of the commutative gauge f\/ields and gauge transformations.
Through this procedure, deformed gauge f\/ield theories can be def\/ined for arbitrary gauge groups/representations.
Consequently, building semi-realistic deformed particle physics models are made much easier.

It is reasonable to expect that the new underlying mathematical structures in the NC gauge f\/ield theories (NCGFT) could lead to
profound observable consequences for the low energy physics.
This is realized by the perturbative loop computation f\/irst proposed by Filk~\cite{Filk:1996dm}.
There are also famous examples of running of the coupling constant in the U(1) NCGFT in the $\star$-product
formalism~\cite{Martin:1999aq}, and the exhibition of fascinating dynamics due to the celebrated ultraviolet/infrared (UV/IR)
phenomenon, without~\cite{Matusis:2000jf, Minwalla:1999px}, and with the Seiberg--Witten
map~\cite{Horvat:2011qg,Horvat:2011bs,Horvat:2013rga,Mehen:2000vs,Schupp:2008fs} included.
Precisely, in~\cite{Bigatti:1999iz,Minwalla:1999px,Matusis:2000jf} it was shown for the f\/irst time how UV short distance
ef\/fects, considered to be irrelevant, could alter the IR dynamics, thus becoming known as the UV/IR mixing.
Some signif\/icant progress on UV/IR mixing and related issues has been
achieved~\cite{Blaschke:2010ck,Blaschke:2009aw,Grosse:2004yu,Magnen:2008pd,Meljanac:2011cs} while a~proper understanding of loop
corrections is still sought for.

More serious ef\/forts on formulating NCQFT models with potential phenomenological inf\/luen\-ce have started for about a~decade ago.
Strong boost came from the Seiberg--Witten map~\cite{Seiberg:1999vs} based enveloping algebra approach, which enables a~direct
deformation of comprehensive phenomenological models like the standard model or
GUTs~\cite{Aschieri:2002mc,Behr:2002wx, Calmet:2001na}.
It appeared then relevant to study ordinary gauge theories with the additional couplings inspired by the SW map/de\-for\-ma\-tion
included~\cite{Aschieri:2002mc,Behr:2002wx, Calmet:2001na}.

To include a~reasonably relevant part of all SW map inspired couplings, one usually calls for an expansion and cut-of\/f
procedure, that is, an expansion of the action in powers of
$\theta^{\mu\nu}$~\cite{Aschieri:2002mc,Behr:2002wx,Calmet:2001na,Hinchliffe:2002km,Martin:2013gma,Trampetic:2008bk}.
Next follows theoretical studies of one loop quantum
properties~\cite{Banerjee:2001un,Bichl:2001cq,Bichl:2001nf,Buric:2013nja,Buric:2007ix,Buric:2010wd,Buric:2006wm,Grimstrup:2002af,Latas:2007eu,
Martin:2002nr,Martin:2009sg,Martin:2009vg},
as well as studies of some new physical phenomena, like breaking of Landau--Yang
theorem,~\cite{Alboteanu:2006hh,Alboteanu:2007bp,Alboteanu:2007by,Buric:2007qx,Minkowski:2003jg,Ohl:2004tn, Schupp:2002up}, etc.
It was also observed that allowing a~deformation-freedom via varying the ratio between individual gauge invariant terms could
improve the renormalizability at one loop level~\cite{Buric:2006wm,Latas:2007eu}.

The studies on phenomenology (possible experimental signal/bounds on noncommutative background) started parallel to the pure
theoretical developments of NCQFTs.
The majority of the accelerator processes had been surveyed up to the second power in $\theta^{\mu
\nu}$~\cite{Alboteanu:2006hh,Alboteanu:2007bp,Alboteanu:2007by, Ohl:2004tn}.
The processes involving photons in noncommutative U(1) gauge theory now involve corrections to the known processes, since the
new couplings, of which the most distinctive being the various photon self couplings, now emerge in the noncommutative
background even at tree level.
Such couplings might give rise to novel processes (normally forbidden in the standard theory) or to provide new channels in the
already known processes.

The formulation of the SW-mapped actions has recently been made exact with respect to the noncommutative parameter $\theta^{\mu
\nu}$~\cite{Mehen:2000vs,Schupp:2008fs,Zeiner:2007}, of\/fering thus an opportunity to compute various processes across full
energy scale~\cite{Horvat:2011iv, Horvat:2010sr}.
Accordingly, one should no longer rely on the expansion in powers in~$\theta$, which could be especially benef\/icial in case when
the quantum gravity scale is not so tantalizing close to the Planck scale.
Thus, in this and several prior work(s) we formulate the $\theta$-exact model action employing formal powers of
f\/ields~\cite{Barnich:2003wq,Grosse:2008dk,Horvat:2012vn,Horvat:2011iv,Horvat:2010sr,Jurco:2001my, Mehen:2000vs,Okawa:2001mv, Zeiner:2007},
aiming at the same time to keep the nonlocal nature of the modif\/ied theory.
Introduction of a~nonstandard momentum dependent quantity of the type $\sin^2({p\theta k}/2)/(p\theta k/2)^2$ into the loop
integrals makes these theories drastically dif\/ferent from their $\theta$-expanded cousins, being thus interesting for pure f\/ield
theoretical reasons.
The deformation-freedom parameters (ratios of weight-parameters of each gauge invariant terms in the actions) are found to be
compatible with the $\theta$-exact action therefore included to study their possible ef\/fects on divergence cancelation(s).

In this review we present closed forms for fermion-loop and photon-loop corrections to the photon and the neutrino two-point
functions using dimensional regularization technique and we combine parameterizations of Schwinger, Feynman, and modif\/ied heavy
quark ef\/fective theo\-ry parameterization (HQET)~\cite{Grozin:2000cm}.
Both two-point functions were obtained as a~function of unspecif\/ied number of the integration dimensions~$D$.
Next we specify gauge f\/ield theory dimension $4$, and discuss the limits $D\to 4$.

The review is structured as follows:
In the following section we describe generalized deformation freedom induced actions, and
we give the relevant Feynman rules.
Sections~\ref{Section3} and~\ref{Section4} are devoted to the computation/presentation/discussions
of photon and neutrino two-point functions containing
the fermion and the photon loop.
Section~\ref{Section5} is devoted to discussions and conclusions.

\section{The model construction}

The main principle that we are implementing in the construction of our $\theta$-exact noncommutative model is that electrically
neutral matter f\/ields will be promoted via hybrid SW map deformations~\cite{Horvat:2011qn} to the neutral noncommutative f\/ields
that couple to photons and transform in the adjoint representation of $\rm U_{\star}(1)$.
We consider a~$\rm U(1)$ gauge theory with a~neutral fermion which decouples from the gauge boson in the commutative limit.
We specify the action and deformation as a~{\it minimal} $\theta$-exact completion of the prior f\/irst order in $\theta$
models~\cite{Behr:2002wx,Buric:2006wm,Calmet:2001na,Minkowski:2003jg,Schupp:2002up}, i.e.\
the new (inter-)action has the prior tri-particle vertices as the leading order.

In the tree-level neutrino-photon coupling processes only vertices of the form $\bar\psi a\psi$ contribute, therefore an
expansion to lowest nontrivial order in $a_\mu$ (but all orders in $\theta$) is enough.
There are at least three known methods for $\theta$-exact computations: The closed formula derived using deformation
quantization based on Kontsevich formality maps~\cite{Jurco:2001my}, the relationship between open Wilson lines in the
commutative and noncommutative picture~\cite{Mehen:2000vs,Okawa:2001mv}, and direct recursive computations using consistency
conditions.
For the lowest nontrivial order a~direct deduction from the recursion and consistency relations
\begin{gather*}
\delta_\Lambda A_\mu=\partial_\mu\Lambda-i[A_\mu \stackrel{\star}{,}\Lambda] \equiv A_\mu[a_\mu+\delta_\lambda
a_\mu]-A_\mu[a_\mu],
\\
\delta_\Lambda\Psi = i[\Lambda \stackrel{\star}{,}\Psi] \equiv\Psi[a_\mu+\delta_\lambda a_\mu, \psi+\delta_\lambda\psi]
-\Psi[a_\mu,\psi],
\\
\Lambda[[\lambda_1,\lambda_2],a_\mu] =[\Lambda[\lambda_1,a_\mu]\stackrel{\star}{,} \Lambda[\lambda_2,a_\mu]]
+i\delta_{\lambda_1}\Lambda[\lambda_2,a_\mu]-i\delta_{\lambda_2} \Lambda[\lambda_1,a_\mu],
\end{gather*}
with the ansatz
\begin{gather*}
\Lambda=\hat\Lambda[a_\mu]\lambda=\big(1+\hat\Lambda^1[a_\mu] +\hat\Lambda^2[a_\mu]+\mathcal O\big(a^3\big)\big)\lambda,
\\
\Psi=\hat\Psi[a_\mu]\psi=\big(1+\hat\Psi^1[a_\mu]+\hat\Psi^2[a_\mu]+\mathcal O\big(a^3\big)\big)\psi,
\end{gather*}
is already suf\/f\/icient.
Capital letters denote noncommutative objects, small letters denote commutative objects, hatted capital letters denote
dif\/ferential operator maps from the latter to the former.
In particular, here $\hat\Psi[a_\mu]$ and $\hat\Lambda[a_\mu]$ are gauge-f\/ield dependent dif\/ferential operators that we shall
now determine: Starting with the fermion f\/ield $\Psi$, at lowest order we have
\begin{gather*}
i[\lambda\stackrel{\star}{,}\psi]=\hat\Psi[\partial\lambda]\psi.
\end{gather*}
Writing the $\star$-commutator explicitly as
\begin{gather*}
[\phi\stackrel{\star}{,}\psi] =\phi(x)\left(e^{i\frac{\partial_x\theta\partial_y}{2}}-e^{-i\frac{\partial_x\theta\partial_y}{2}}\right)\psi(y)\bigg|_{x=y}
=2i\phi(x)\sin\left(\frac{\partial_x\theta\partial_y}{2}\right)\psi(y)\bigg|_{x=y}
\\
\phantom{[\phi\stackrel{\star}{,}\psi]}
=i\theta^{ij}\bigg(\frac{\partial \phi(x)}{\partial x^i}\bigg) \frac{\sin\big(\frac{\partial_x\theta\partial_y}{2}\big)}
{\frac{\partial_x\theta\partial_y}{2}} \bigg(\frac{\partial \psi(y)}{\partial y^{j}}\bigg) \bigg|_{x=y},
\end{gather*}
we observe that
\begin{gather*}
\hat\Psi[a_\mu]=-\theta^{ij}a_i\star_2\partial_j,
\end{gather*}
will fulf\/ill the consistency relation.
The generalized star-product $\star_2$, appearing in the above, is def\/ined, respectively,
as~\cite{Horvat:2011qg,Horvat:2011bs, Mehen:2000vs,Schupp:2008fs}:
\begin{gather*}
\phi(x)\star_2 \psi(x)=\frac{\sin\frac{\partial_1\theta \partial_2}{2}}{\frac{\partial_1\theta
\partial_2}{2}}\phi(x_1)\psi(x_2)\bigg|_{x_1=x_2=x}.
\end{gather*}
Here $\star$-product~\eqref{f*g} is associative but noncommutative, while $\star_2$ is commutative but nonassociative.
The $\star$-commutator can then be rewritten as the following $\star_2$-products
\begin{gather}
\label{rel}
[\phi\stackrel{\star}{,}\psi]=i\theta^{ij}\partial_i\phi\star_2\partial_j\psi.
\end{gather}

The gauge transformation $\Lambda$ can be worked out similarly, namely
\begin{gather*}
0=[\lambda_1\stackrel{\star}{,}\lambda_2] +i\hat\Lambda[\partial\lambda_1]\lambda_2 -i\hat\Lambda[\partial\lambda_2]\lambda_1
=\frac{1}{2}\big([\lambda_1\stackrel{\star}{,}\lambda_2] -[\lambda_2\stackrel{\star}{,}\lambda_1]\big)
+i\hat\Lambda[\partial\lambda_1]\lambda_2 -i\hat\Lambda[\partial\lambda_2]\lambda_1,
\end{gather*}
and hence
\begin{gather*}
\hat\Lambda^1=-\frac{1}{2}\theta^{ij}a_i\star_2\partial_j.
\end{gather*}
The gauge f\/ield $a_\mu$ requires slightly more work.
The lowest order terms in its consistency relation are
\begin{gather*}
-\partial_\mu\left(\frac{1}{2}\theta^{ij}a_i\star_2\partial_j \lambda\right)-i[\lambda\stackrel{\star}{,}a_\mu]
=A^2_\mu[a_\mu+\partial_\mu\lambda]-A^2_\mu[a_\mu],
\end{gather*}
where $A^2$ is the $a^2$ order term in the expansion of~$A$ as power series of~$a$.
Using the relation~\eqref{rel}, the left hand side can be rewritten as $-\frac{1}{2}\theta^{ij}\partial_\mu
a_i\star_2\partial_j\lambda-\frac{1}{2} \theta^{ij}a_i\star_2\partial_\mu\partial_j\lambda
-\theta^{ij}\partial_i\lambda\star_2\partial_j a_\mu$, where the f\/irst term comes from $-\frac{1}{2}\theta^{ij}\partial_\mu
a_i\star_2 a_j$, while the third one comes from $-\theta^{ij}a_i\star_2\partial_j a_\mu$.
After a~gauge transformation, the sum of the f\/irst and third terms equals the second term.
Ultimately, we obtain SW map solutions up to the $\mathcal{O}\big(a^2\big)$ order:
\begin{gather}
A_\mu =a_\mu-\frac{1}{2}\theta^{\nu\rho}{a_\nu}\star_2(\partial_\rho a_\mu+f_{\rho\mu})+\mathcal O\big(a^3\big),
\nonumber
\\
\Psi =\psi-\theta^{\mu\nu} {a_\mu}\star_2{\partial_\nu}\psi+\mathcal O\big(a^2\big)\psi,
\qquad
\Lambda =\lambda-\frac{1}{2}\theta^{\mu\nu}a_\mu\star_2\partial_\nu\lambda+\mathcal O\big(a^2\big)\lambda,
\label{SWmap}
\end{gather}
with $f_{\mu\nu}$ being the commutative Abelian f\/ield strength $f_{\mu\nu}=\partial_\mu a_\nu-\partial_\nu a_\mu$.

The resulting expansion def\/ines in the next section the one-photon-two-fermion and the three-photon vertices, $\theta$-exactly.

\subsection{Actions}

We start with the {\it minimal} NC model of a~SW type $\rm U_{\star}(1)$ gauge theory on Euclidean space\-time.
Here in the starting action, the {\it minimal} refers on the number of f\/ields/gauge f\/ield strengths/covariant derivatives: two
gauge f\/ields, two gauge f\/ield strengths, one covariant derivative and three f\/ields for gauge-fermion interactions.
Thus we have
\begin{gather}
S^{\rm min}=\int-\frac{1}{2}F^{\mu\nu}\star F_{\mu\nu}+i\bar\Psi\star\fmslash{D}\Psi,
\label{S}
\end{gather}
with the coupling constant to be set as $e=1$, and with the following def\/initions of the non-Abelian NC covariant derivative and
the f\/ield strength, respectively:
\begin{gather*}
D_\mu\Psi=\partial_\mu\Psi-i[A_\mu\stackrel{\star}{,}\Psi]
\qquad
\mbox{and}
\qquad
F_{\mu\nu}=\partial_\mu A_\nu-\partial_\nu A_\mu-i[A_\mu\stackrel{\star}{,}A_\nu].
\end{gather*}
All the f\/ields in this action are images under (hybrid) Seiberg--Witten maps of the corresponding commutative f\/ields $a_\mu$ and
$\psi$.
In the original work of Seiberg and Witten and in virtually all subsequent applications, these maps are understood as (formal)
series in powers of the noncommutativity parameter $\theta^{\mu\nu}$.
Physically, this corresponds to an expansion in momenta and is valid only for low energy phenomena.
Here we shall not subscribe to this point of view and instead interpret the noncommutative f\/ields as valued in the enveloping
algebra of the underlying gauge group.
This naturally corresponds to an expansion in powers of the gauge f\/ield $a_\mu$ and hence in powers of the coupling
constant~$e$.
At each order in $a_\mu$ we shall determine $\theta$-exact expressions.
In the following we discuss the model construction for the photon and the massless fermion case.
Since we have set $e=1$, to restore the coupling constant one simply substitutes $a_\mu$ by $e a_\mu$ and then divides the
gauge-f\/ield term in the Lagrangian by $e^2$.
Coupling constant~$e$, carries (mass) dimension $(4-d)/2$ in the~$d$-dimensional f\/ield theory.

The expansion in powers of the commutative gauge f\/ield content is motivated by the obvious fact that in perturbative quantum
f\/ield theory one can sort the interaction vertices by the number of external legs and this is equivalent to the number of f\/ield
operators in the corresponding interacting terms.
For any specif\/ic process and loop order there exists an upper limit on the number of external legs.
So if one expands the noncommutative f\/ields with respect to the formal power of the commutative f\/ields which are the primary
f\/ields in the theory up to an appropriate order, the relevant vertices in a~specif\/ic diagram will automatically be exact to all
orders of $\theta$.

The {\it minimal} gauge invariant nonlocal interaction~\eqref{S} includes the gauge boson self-coupling as well as the
fermion-gauge boson coupling, denoted here as $S_{\rm g}$ and $S_{\rm f}$, respectively:
\begin{gather*}
S^{\rm min}=S_{\rm U(1)}+S_{\rm g}+S_{\rm f}.
\end{gather*}

In the next step we expand the action~\eqref{S} in terms of the commutative gauge parameter $\lambda$ and f\/ields $a_\mu$ and
$\psi$ using the $\rm U(1)$ SW map solutions~\eqref{SWmap}.
This way, the photon self-interaction up to the lowest nontrivial order is obtained~\cite{Horvat:2011qg,Horvat:2013rga}:
\begin{gather}
S_g= \int if^{\mu\nu}\star[a_\mu\stackrel{\star}{,}a_\nu] +\partial_\mu
\big(\theta^{\rho\sigma}a_\rho\star_2(\partial_\sigma a_{\nu}+f_{\sigma\nu})\big)\star f^{\mu\nu}+\mathcal O\big(a^4\big)
\nonumber
\\
\phantom{S_g}
= \int\theta^{\rho\tau}f^{\mu\nu}\left(\frac{1}{4}f_{\rho\tau}\star_2f_{\mu\nu}-f_{\mu\rho}\star_2 f_{\nu\tau}\right)+\mathcal
O\big(a^4\big).
\label{Sgauge}
\end{gather}
The the lowest order photon-fermion interaction (f\/irst three terms of equation~(2.7) from~\cite{Horvat:2011qg}) reads as follows
\begin{gather}
S_f= \int \bar\psi\gamma^\mu[a_\mu\stackrel{\star}{,}\psi] +i(\theta^{ij}\partial_i\bar\psi \star_2
a_j)\fmslash\partial\psi-i\bar\psi\star \fmslash\partial(\theta^{ij} a_i\star_2\partial_j\psi)+\bar\psi\mathcal{O}\big(a^2\big)\psi
\nonumber
\\
\phantom{S_f}
= -\int i\theta^{\rho\tau}\bar\psi\gamma^\mu\left(\frac{1}{2}f_{\rho\tau}\star_2\partial_\mu\psi-
f_{\mu\rho}\star_2\partial_\tau\psi\right)+\bar\psi\mathcal{O}\big(a^2\big)\psi.
\label{Sfermion}
\end{gather}
Note that actions for the gauge and the matter f\/ields obtained above,~\eqref{Sgauge} and~\eqref{Sfermion} respectively, are
nonlocal objects due to the presence of the (generalized) star products.

\subsection[General deformed actions: $S_{\rm f}$ and $S_{\rm g}$]{General deformed actions:
$\boldsymbol{S_{\rm f}}$ and $\boldsymbol{S_{\rm g}}$}
\label{ss2.2}

It is easy to see that each of the interactions~\eqref{Sgauge} and~\eqref{Sfermion} contains two U(1) gauge invariant terms,
therefore one could vary the ratio between them without spoiling the gauge invariance.
Prior studies have also indicated that varying these ratios can improve the one-loop behavior of the
model~\cite{Buric:2006wm,Horvat:2011qg,Horvat:2013rga,Horvat:2011iv,Latas:2007eu}.
For this propose we introduce further two-dimensional deformation-parameter-space $(\kappa_f,\kappa_g)$.

The deformation parameter $\kappa_f$ in the photon-gauge boson interaction can be so chosen that it realizes the linear
superposition of two possible nontrivial noncommutative deformations of a~free neutral fermion action proposed
in~\cite{Horvat:2011qg,Horvat:2013rga,Horvat:2011iv}.
Its existence was already hinted in the $\theta$-expanded expressions in~\cite{Schupp:2002up} but not fully exploited in the
corresponding loop computation before.

The pure gauge action $S_{\rm g}$ deformation $\kappa_g$ was f\/irst presented in the non-Abelian gauge sector action of the NCSM
and NC SU(N) at f\/irst order in $\theta$, $S_{\rm g}^\theta$~\cite{Buric:2006wm,Latas:2007eu}.
This could be realized by generalizing the standard SW map expression for linear in $\theta$ gauge f\/ield strength into the
following form~\cite{Trampetic:2007ze}:
\begin{gather*}
F_{\mu\nu}^{\theta}(\kappa_g) = f_{\mu\nu} + \theta^{\rho\tau}\big(\kappa_g^{-1}f_{\mu\rho}f_{\nu\tau}-a_{\rho}\partial_{\tau}
f_{\mu\nu}\big)+\mathcal{O}\big(\theta^2\big).
\end{gather*}
The gauge transformation for the noncommutative f\/ield strength
$\delta_{\lambda}F_{\mu\nu}=i[\Lambda\stackrel{\star}{,}F_{\mu\nu}]$ will still be satisf\/ied at its leading order in $\theta$.
We have observed in prior studies~\cite{Horvat:2012vn,Horvat:2013rga,Trampetic:2007ze} that the above deformation can be made
$\theta$-exact and adopted it here
\begin{gather}
F_{\mu\nu}(\kappa_g) = f_{\mu\nu} + \theta^{\rho\tau}
\big(\kappa_g^{-1}f_{\mu\rho}\star_{2}f_{\nu\tau}-a_{\rho}\star_2\partial_{\tau} f_{\mu\nu}\big)+\mathcal{O}\big(a^3\big).
\label{fieldsa2}
\end{gather}
Using the relation~\eqref{rel} we can see that $\delta_\lambda
F_{\mu\nu}(\kappa_g)=i[\lambda\stackrel{\star}{,}f_{\mu\nu}]+\mathcal{O}\big(a^2\big)\lambda$, which represents again the desired f\/ield
strength consistency at the corresponding order.
Thus, starting with~\eqref{S},~\eqref{fieldsa2} and~\eqref{Sfermion}, followed by an appropriate f\/ield strength redef\/inition
$f_{\mu\nu}\to\kappa_g f_{\mu\nu}$, and f\/inally after an overall rescaling $\kappa_g^{-2}$, we can write the generalized
manifestly gauge invariant actions with {\it minimal} number of f\/ields\footnote{It could be simpler if we have associated
$\kappa_g$ with $f_{\mu\rho}$ in~\eqref{g} as the $\kappa_f$ in~\eqref{f}, we choose the other way around to unify our result
with the prior works~\cite{Buric:2006wm,Horvat:2012vn,Horvat:2013rga,Latas:2007eu,Trampetic:2007ze}.}:
\begin{gather}
S_{\rm U(1)}=\int-\frac{1}{2}f_{\mu\nu}f^{\mu\nu}+i\bar\psi\fmslash\partial\psi,
\\
S_{\rm g}(\kappa_g)=\int\theta^{\rho\tau}f^{\mu\nu}\left(\frac{\kappa_g}{4}f_{\rho\tau}\star_2f_{\mu\nu}-f_{\mu\rho}\star_2
f_{\nu\tau}\right),
\label{g}
\\
S_{\rm f}(\kappa_f)=-\int i\theta^{\rho\tau}\bar\psi\gamma^\mu\left(\frac{1}{2}f_{\rho\tau}\star_2\partial_\mu\psi-\kappa_f
f_{\mu\rho}\star_2\partial_\tau\psi\right).
\label{f}
\end{gather}

Since $S_{\rm g}(\kappa_g)$ and $S_{\rm f}(\kappa_f)$ are both gauge invariant by themselves, one can incorporate either one or
both of them into the full Lagrangian.
The above actions were obtained by a~$\theta$-exact gauge-invariant truncation of a~$\rm U_\star(1)$ model up to tri-leg
vertices.
Such an operation is achievable because the $\rm U(1)$ gauge transformation after deformation preserves the number of f\/ields
within each term.

Motivation to introduce deformation parameters $\kappa_g$ and $\kappa_f$ was, besides the general gauge invariance of the
action, to help eliminating one-loop pathologies due to the UV and/or IR divergences in both sectors.
The parameter-space $(\kappa_f,\kappa_g)$ represents a~measure of the deformation-freedom in the matter $S_{\rm f}(\kappa_f)$
and the gauge $S_{\rm g}(\kappa_g)$ sectors, respectively.
We should clarify, that we are interested in the general gauge invariant interactions induced by the $\theta^{\mu\nu}$
background instead of the strictly Moyal--Weyl star-product deformation of the commutative gauge theories and its Seiberg--Witten
map extension.
We relax the constraint that a~deformation should be Moyal--Weyl type for the hope that such variation could provide certain
additional control on the novel pathologies due to the noncommutativity, which had indeed occurred in the $\theta$-expanded
models studied before, and as we will discuss later, in our $\theta$-exact model as well.
We still constrained our model building by requiring that {\em each} of the gauge invariant interaction terms arises within
a~Seiberg--Witten map type deformation, only their linear combination ratios $\kappa_g$ and $\kappa_f$ are allowed to vary.
This is all explained in full details in~\cite{Horvat:2011qg,Horvat:2013rga,Horvat:2011iv}.
Each parameter bears the origin from the corresponding $\theta$-expanded theory~\cite{Buric:2006wm,Latas:2007eu,Schupp:2002up}.

By straightforward reading-out procedure from $S_{\rm g}$~\eqref{g} we obtain the following Feynman rule for the triple-photon
vertex in momentum space:
\begin{gather}
\Gamma_{\kappa_g}^{\mu\nu\rho}(p,k,q)=F(k,q)V_{\kappa_g}^{\mu\nu\rho}(p,k,q),
\qquad
F(k,q)=\frac{\sin\frac{k\theta q}{2}}{\frac{k\theta q}{2}},
\label{Fg}
\end{gather}
where momenta $p$, $k$, $q$ are taken to be incoming satisfying the momentum conservation $p+k+q$ $=0$~\cite{Horvat:2013rga}.
The deformation freedom ambiguity~$\kappa_g$ is included in the vertex function:
\begin{gather}
V_{\kappa_g}^{\mu\nu\rho} (p,k,q)=-(p\theta k)[(p-k)^{\rho}g^{\mu\nu}+(k-q)^\mu g^{\nu\rho}+(q-p)^{\nu}g^{\mu\rho}]
\nonumber
\\
\hphantom{V_{\kappa_g}^{\mu\nu\rho} (p,k,q)=}
{}-\theta^{\mu\nu}[p^{\rho}(k q)-k^{\rho}(p q)]-\theta^{\nu\rho} [k^{\mu}(p q)-q^{\mu}(p k)]
-\theta^{\rho\mu}[q^{\nu}(p k)-p^{\nu}(k q)]
\nonumber
\\
\hphantom{V_{\kappa_g}^{\mu\nu\rho} (p,k,q)=}
{}+(\theta p)^\nu[g^{\mu\rho}q^2-q^\nu q^{\rho}] +(\theta p)^{\rho}[g^{\mu\nu}k^2-k^\mu k^\nu]+(\theta
k)^\mu[g^{\nu\rho}q^2-q^\nu q^{\rho}]
\nonumber
\\
\hphantom{V_{\kappa_g}^{\mu\nu\rho} (p,k,q)=}
{}+(\theta k)^{\rho}[g^{\mu\nu}p^2-p^\mu p^\nu]+(\theta q)^\nu[g^{\mu\rho}p^2-p^\mu p^{\rho}] +(\theta
q)^\mu[g^{\nu\rho}k^2-k^\nu k^{\rho}]
\nonumber
\\
\hphantom{V_{\kappa_g}^{\mu\nu\rho} (p,k,q)=}
{}+(\kappa_g-1)\big((\theta p)^{\mu}[g^{\nu\rho}(k q)-q^\nu k^{\rho}]+ (\theta k)^{\nu}[g^{\mu\rho}(q p)-q^\mu
p^{\rho}]
\nonumber
\\
\hphantom{V_{\kappa_g}^{\mu\nu\rho} (p,k,q)=}
{}+(\theta q)^{\rho} [g^{\mu\nu}(k p)-k^\mu p^{\nu}]\big).
\label{Fga}
\end{gather}
The above vertex function~\eqref{Fga} is in accord with corresponding Feynman rule for triple neutral gauge-boson coupling
in~\cite{Buric:2007qx}.

From $S_{\rm f}$~\eqref{f} the fermion-photon vertex reads as follows
\begin{gather}
\Gamma_{\kappa_f}^\mu(k,q)=F(k,q) V_{\kappa_f}^\mu(k,q)
=F(k,q)\big[\kappa_f\big(\fmslash{k}(\theta q)^\mu-\gamma^\mu(k\theta q)\big)-(\theta k)^\mu\fmslash{q}\big],
\label{Ff}
\end{gather}
where~$k$ is the photon incoming momentum, and the fermion momentum~$q$ f\/lows through the vertex, as it should~\cite{Horvat:2013rga}.

\section{Photon two-point function}\label{Section3}

\subsection{Computing photon two-point function using dimensional regularization}

Employing the parametrization given in this section we illustrate the way we have performed the computation of the integrals
which dif\/fer from regular ones by the existence of a~non-quadratic $k\theta p$ denominators.
The key point was to introduce the HQET parametrization~\cite{Grozin:2000cm}, represented as follows
\begin{gather*}
\frac{1}{a_1^{n_1} a_2^{n_2}}= \frac{\Gamma(n_1+n_2)}{\Gamma(n_1)\Gamma(n_2)} \int_0^\infty\frac{i^{n_1}y^{n_1-1} dy}{(ia_1y +
a_2)^{n_1+n_2}}.
\end{gather*}
To perform computations of our integrals, we f\/irst use the Feynman parametrization on the quadratic denominators, then the HQET
parametrization help us to combine the quadratic and linear denominators.
For example
\begin{gather*}
\frac{1}{k^2(p+k)^2}\frac{1}{k\theta p} = 2i\int_0^1 dx\int_0^\infty dy
\big[\big(k^2+i\epsilon\big)(1-x)+\big((p+k)^2+i\epsilon\big)x+iy(k\theta p)\big]^{-3}.
\end{gather*}
After employing the Schwinger parametrization, the phase factors from~\eqref{Ff} can be absorbed by redef\/ining the~$y$ integral.
This way we obtain
\begin{gather*}
\frac{2-e^{ik\theta p}-e^{-ik\theta p}}{k^2(p+k)^2(k\theta p)}\cdot\{\rm numerator\}
\\
\qquad
=2i\int_0^1\! dx\int_0^{\frac{1}{\lambda}}\! dy\int_0^\infty \! d\lambda\lambda^2
e^{-\lambda\big(l^2+x(1-x)p^2+\frac{y^2}{4}(\theta p)^2\big)}
\cdot\{y\ \text{odd~terms~of~the~numerator}\},
\end{gather*}
with loop-momenta being $l=k+xp+\frac{i}{2}y(\theta p)$.
By this means the~$y$-integral limits take the places of planar/nonplanar parts of the loop integral.
For higher negative power(s) of $k\theta p$, the parametrization follows the same way except the appearance of the
additional~$y$-integrals which lead to {\em finite} hypergeometric functions\footnote{See \url{http://functions.wolfram.com/07.32.06.0031.01}.}.
Following~\cite{Horvat:2011qn}, we are enabled to follow the general procedure of dimensional regularization in computing
one-loop two point functions.
Thus we start the computations with respect to general integration dimension~$D$, next we set the $D\to 4$ limits and perform
the full analysis of the one-loop two point functions behavior.

\begin{figure}[t]\centering
\includegraphics[width=5.5cm]{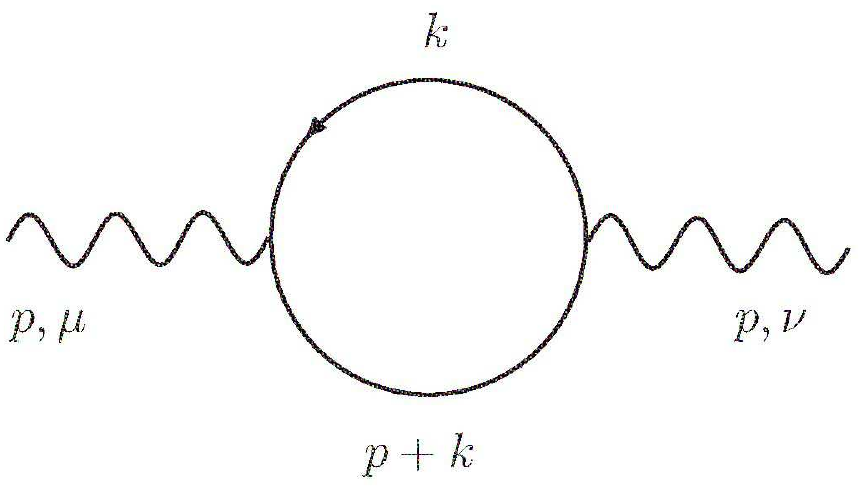}
\caption{Fermion-loop contribution to the photon two-point function.}
\label{FigFydiagr1}
\end{figure}

{\bf Photon two-point function: fermion-loop.}
The fermion-loop contribution is read out from Fig.~\ref{FigFydiagr1}.
Dimensional regularization than gives photon polarization tensor full expression
\begin{gather*}
\Pi_{\kappa_f}^{\mu\nu}(p)_D=-\tr\mu^{4-D}\int\frac{d^D k}{(2\pi)^D}\Gamma_{\kappa_f}^\mu(-p,p+k)\frac{i(\fmslash p+\fmslash
k)}{(p+k)^2}\Gamma_{\kappa_f}^\nu(p,k)\frac{i\fmslash k}{k^2},
\end{gather*}
where the momentum structure and dependence on the parameter $\kappa_f$ is encoded.
Performing a~large amount of computations we have found the following structure:
\begin{gather}
\Pi_{\kappa_f}^{\mu\nu}(p)_D =\frac{1}{(4\pi)^2}\big[\big(g^{\mu\nu}p^2-p^\mu p^\nu\big)F_1^{\kappa_f}(p) + (\theta
p)^\mu(\theta p)^\nu F_{2}^{\kappa_f}(p) \big],
\label{Flfinal}
\end{gather}
while the full details of the loop-coef\/f\/icients $F_{1,2}^{\kappa_f}(p)$ computations are given in~\cite{Horvat:2013rga}.
It is straightforward to see that each term in the tensor structure~\eqref{Flfinal} does satisfy the Ward identity by itself,
therefore $ p_\mu\Pi_{\kappa_f}^{\mu\nu}(p)_D=p_\nu\Pi_{\kappa_f}^{\mu\nu}(p)_D=0$, $\forall\, (D,\kappa_f)$.

In the limit $D\to 4-\epsilon$, the loop-coef\/f\/icients can be expressed in the following closed forms:
\begin{gather}
F_1^{\kappa_f}(p) =  -\kappa_f^2\frac{8}{3}\bigg[\frac{2}{\epsilon}+\ln{\pi e^{\gamma_{\rm E}}}+\ln\big(\mu^2(\theta p)^2\big)\bigg]
+4\kappa_f^2p^2(\theta p)^2\sum\limits_{k=0}^\infty\frac{(k+2)(p^2(\theta p)^2)^k}{4^k\Gamma[2k+6]}
\nonumber
\\
\phantom{F_1^{\kappa_f}(p) =}{}
\times
\big[(k+2)\big(\ln \big(p^2(\theta p)^2\big)-\psi(2k+6)-\ln4\big)+2\big],
\label{FlfinalF1}
\\
F_{2}^{\kappa_f}(p) =  \kappa_f\frac{8}{3}\frac{p^2}{(\theta p)^2}\bigg[\kappa_f-8\big(\kappa_f+2\big)\frac{1}{p^2(\theta p)^2}\bigg]
-4\kappa_f^2p^4\sum\limits_{k=0}^\infty\frac{(p^2(\theta p)^2)^k}{4^k\Gamma[2k+6]}
\nonumber
\\
\phantom{F_1^{\kappa_f}(p) =}{}
\times
\big[(k+1)(k+2)\big(\ln\big(p^2(\theta p)^2\big)-2\psi(2k+6)-\ln4\big)+2k+3\big],
\label{FlfinalF2}
\end{gather}
with $\gamma_{\rm E}\simeq0.577216$ being Euler's constant.
The above expressions for $F_{1,2}^{\kappa_f}(p)$ contain both contributions, from the planar as well as from the non-planar
graphs.
All of the {\it divergences} arising from the fermion-loop (Fig.~\ref{FigFydiagr1}) could be removed by the unique choice
$\kappa_f=0$, as in that case the whole general amplitude~\eqref{Flfinal} vanishes for any integration dimensions
D~\cite{Horvat:2013rga}.

Evaluation of the four-dimensional $\theta$-exact fermion-part contribution to the photon pola\-ri\-zation tensor, i.e.\
the fermion-loop photon two-point function~\eqref{Flfinal} yields two already known tensor
structures~\cite{Brandt:2001ud, Hayakawa:1999yt,Hayakawa:1999zf}.
The loop-coef\/f\/icients $F^{\kappa_f}_{1,2}(p)$, on the other hand, exhibit nontrivial $\kappa_f$ dependence.
Namely, in the limit $\kappa_f \to 0$ $\Longrightarrow$ $ F^{\kappa_f}_1(p)=F^{\kappa_f}_2(p)=0$, thus the photon polarization
tensor~\eqref{Flfinal} vanishes, while $\kappa_f=1$ appears to be identical to the non SW-map model.
Fermion-loop contains UV and logarithmic divergence in $F^{\kappa_f}_1(p)$ for $\kappa_f\neq 0$, while the quadratic UV/IR
mixing could be removed by setting $\kappa_f =0,-2$ in $F^{\kappa_f}_2(p)$.

Finally, it is important to stress that there is an additional fermion-loop tadpole diagram contribution to the photon 2-point
function, arising from 2-photon-2-fermion ($\bar\psi a^2 \psi$) interaction vertices~\cite{Horvat:2011qg,Horvat:2011iv}.
However, it was shown in~\cite{Schupp:2008fs}, that this tadpole diagram vanishes due to the internal Lorentz structure.

\begin{figure}[t]\centering
\includegraphics[width=5.5cm]{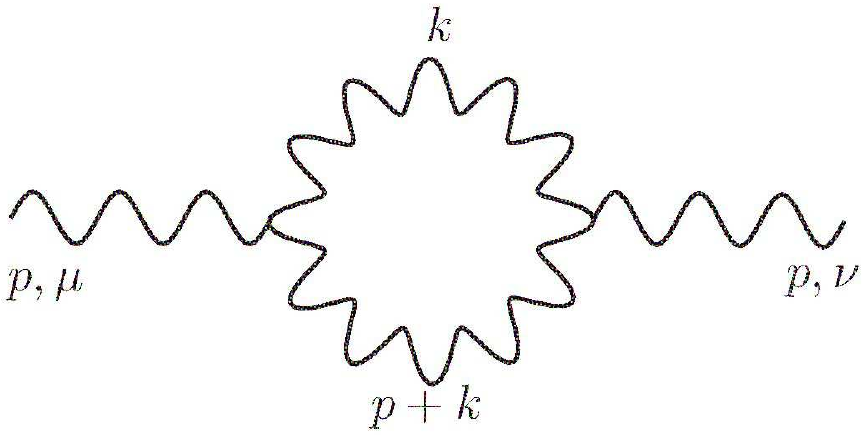}
\caption{Photon-loop contribution to the photon two-point function.}
\label{FigFydiagr2}
\end{figure}

{\bf Photon two-point function: photon-loop.}
The photon-loop computation involves a~single photon-loop integral contribution to the photon polarization tensor from
Fig.~\ref{FigFydiagr2}.
Using dimensional regularization, in~\cite{Horvat:2013rga} we have computed the integral,
\begin{gather*}
\Pi_{\kappa_g}^{\mu\nu}(p)_D = \frac{1}{2}\mu^{4-D}\int\frac{d^D k}{(2\pi)^D} \Gamma_{\kappa_g}^{\mu\rho\sigma}
(-p;-k,p+k)\frac{-ig_{\rho\rho'}}{k^2}\Gamma_{\kappa_g}^{\nu\rho'\sigma'} (p;k,-k-p) \frac{-ig_{\sigma\sigma'}}{(p+k)^2},
\end{gather*}
as a~function of deformation freedom $\kappa_g$ ambiguity.
We obtained the following compact form of the photon-loop contribution to the photon two-point function in D-dimensions,
\begin{gather}
\Pi_{\kappa_g}^{\mu\nu}(p)_D =\frac{1}{(4\pi)^2}\Big\{\big[g^{\mu\nu}p^2-p^\mu p^\nu\big]B_1^{\kappa_g}(p) +(\theta p)^\mu
(\theta p)^\nu B_2^{\kappa_g}(p)
\nonumber
\\
\phantom{\Pi_{\kappa_g}^{\mu\nu}(p)_D =}
{}+\big[g^{\mu\nu}(\theta p)^2-(\theta\theta)^{\mu\nu}p^2 + p^{\{\mu}(\theta\theta p)^{\nu\}}\big]B_3^{\kappa_g}(p)
\nonumber
\\
\phantom{\Pi_{\kappa_g}^{\mu\nu}(p)_D =}
{}+\big[(\theta\theta)^{\mu\nu}(\theta p)^2+(\theta\theta p)^\mu(\theta\theta p)^\nu\big]B_4^{\kappa_g}(p)
+(\theta p)^{\{\mu}(\theta\theta\theta p)^{\nu\}} B_5^{\kappa_g}(p)\Big\}.
\label{PlD}
\end{gather}
Each term of the above tensor structure satisf\/ies Ward identities by itself, i.e.\
$p_\mu\Pi_{\kappa_g}^{\mu\nu}(p)_D=p_\nu\Pi_{\kappa_g}^{\mu\nu}(p)_D=0, \forall (D,\kappa_g)$.

The structure of the photon-loop contribution to the photon polarization tensor contains various previously unknown new momentum
structures.
It is much reacher with respect to earlier non SW map $\theta$-exact results based on $\star$-product
only~\cite{Brandt:2001ud, Hayakawa:1999yt}.
These higher order in $\theta$ ($\theta\theta\theta\theta$ types) terms suggest certain connection to the open/closed string
correspondence~\cite{Jurco:2001my, Seiberg:1999vs} (in an inverted way).
We consider such connection plausible given the connection between noncommutative f\/ield theory and quantum gravity/string
theory.

All coef\/f\/icients $B^{\kappa_g}_i(p)$ can be expressed as sum over integrals over modif\/ied Bessel and generalized hypergeometric
functions.
A complete list of coef\/f\/icients $B^{\kappa_g}_i(p)$ including computations, as a~functions of dimension~$D$, is given in full
details in~\cite{Horvat:2013rga}.
Next we concentrate on the {\em divergent} parts in the limit $D\to 4-\epsilon$, and in the IR regime only
\begin{gather}
B_1^{\kappa_g}(p)\sim \bigg(\frac{2}{3}(\kappa_g-3)^2 +\frac{2}{3}(\kappa_g+2)^2
\frac{p^2(\tr\theta\theta)}{(\theta p)^2} +\frac{4}{3}\big(\kappa^2_g+4\kappa_g+1\big) \frac{p^2(\theta\theta p)^2}{(\theta p)^4}\bigg)
\nonumber
\\
\phantom{B_1^{\kappa_g}(p)\sim}
\times
\left[\frac{2}{\epsilon} + \ln(\mu^2(\theta p)^2)\right]-\frac{16}{3}(\kappa_g-1)^2\frac{1}{(\theta p)^6}
\big((\tr\theta\theta)(\theta p)^2+4(\theta\theta p)^2\big),
\label{PIB1}
\\
B_2^{\kappa_g}(p)\sim \bigg(\frac{8}{3}(\kappa_g-1)^2 \frac{p^4(\theta\theta p)^2}{(\theta
p)^6}+\frac{2}{3}\big(\kappa_g^2-2\kappa_g-5\big)\frac{p^4(\tr\theta\theta)}{(\theta p)^4}+\frac{2}{3}\big(25\kappa^2_g-86\kappa_g+73\big)
\nonumber
\\
\phantom{B_2^{\kappa_g}(p)\sim}
\times
\frac{p^2}{(\theta p)^2}\bigg)\left[\frac{2}{\epsilon} + \ln\big(\mu^2(\theta p)^2\big)\right]
-\frac{16}{3}(\kappa_g-3)(3\kappa_g-1)\frac{1}{(\theta p)^4}
\nonumber
\\
\phantom{B_2^{\kappa_g}(p)\sim}
{}+\frac{32}{3}(\kappa_g-1\big)^2\frac{1}{(\theta p)^8}\big((\tr\theta\theta)(\theta p)^2+6(\theta\theta p)^2\big),
\\
B_3^{\kappa_g}(p)\sim-\frac{1}{3}\big(\kappa_g^2-2\kappa_g-11\big)\frac{p^2}{(\theta p)^2} \left[\frac{2}{\epsilon} +
\ln\big(\mu^2(\theta p)^2\big)\right] -\frac{8}{3(\theta p)^4}\big(\kappa_g^2-10\kappa_g+17\big),
\\
B_4^{\kappa_g}(p)\sim-2(\kappa_g+1)^2\frac{p^4}{(\theta p)^4}\left[\frac{2}{\epsilon} + \ln\big(\mu^2(\theta
p)^2\big)\right]-\frac{32p^2}{3(\theta p)^6}\big(\kappa_g^2-6\kappa_g+7\big),
\\
B_5^{\kappa_g}(p)\sim\frac{4}{3}\big(\kappa_g^2+\kappa_g+4\big)\frac{p^4}{(\theta p)^4} \left[\frac{2}{\epsilon} +
\ln\big(\mu^2(\theta p)^2\big)\right]+\frac{64p^2}{3(\theta p)^6}(\kappa_g-1)(\kappa_g-2).
\label{PIB5}
\end{gather}
Note that all $B_i^{\kappa_g}(p)$ coef\/f\/icients are computed for abitrary $\kappa_g$ and the notation $\sim$ means that in the
above equations we have neglected all f\/inite terms.
We observe here the presence of the UV divergences as well as a~quadratic UV/IR mixing in all $B^{\kappa_g}_i$'s.
Up to the $1/\epsilon$ terms, the UV divergence is at most logarithmic, i.e.\
there is a~logarithmic ultraviolet/infrared term representing a~soft UV/IR mixing.
The results~\eqref{PIB1}--\eqref{PIB5} in four dimensions
for arbitrary $\kappa_g$ show power type UV/IR mixing, therefore diverge
at both the commutative limit ($\theta\to 0$) and the size-of-the-object limit ($|\theta p|\to 0$).
Inspecting~\eqref{PIB1} to~\eqref{PIB5} together with general structure~\eqref{PlD} we found decouplings of UV and logarithmic
IR divergences from the power UV/IR mixing terms.
The latter exists in all $B^{\kappa_g}_i(p)$'s.
The logarithmic IR divergences from planar and nonplanar sources appear to have identical coef\/f\/icient and combine into a~single
$\ln\mu^2(\theta p)^2$ term.
Finally it is important to stress that no single $\kappa_g$ value is capable of removing all novel divergences.

\subsection[Photon-loop with a~special $\theta^{\mu\nu}$ in four dimensions]{Photon-loop with
a~special $\boldsymbol{\theta^{\mu\nu}}$ in four dimensions}

In our prior analysis we have found that in the $D\to 4-\epsilon$ limit the general of\/f-shell contribution of photon
self-interaction loop to the photon two-point function contains complicated non-vanishing UV and IR divergent terms with
existing and new momentum structures, regardless the $\kappa_g$ values we take.
To see whether there exists certain remedy to this situation we explore two conditions which have emerged in the prior studies.
First we tested the zero mass-shell condition/limit ($p^2\to0$) used in $\theta$-expanded models~\cite{Martin:2009vg}.
Inspection of equations~\eqref{FlfinalF1}--\eqref{FlfinalF2} and~\eqref{PIB1}--\eqref{PIB5}
show some simplif\/ication but not the full cancelation of the pathological divergences.
Such condition clearly appears to be unsatisfactory.

Next we have turned into the other one, namely the special full rank $\theta^{\mu\nu}$ choice
\begin{gather}
\theta^{\mu\nu}\equiv \theta^{\mu\nu}_{\sigma_2}=\frac{1}{\Lambda_{\rm NC}^2}
\begin{pmatrix}
0&-1&0&0
\\
1&0&0&0
\\
0&0&0&-1
\\
0&0&1&0
\end{pmatrix}
=\frac{1}{\Lambda_{\rm NC}^2}
\begin{pmatrix}
{i\sigma_2}&0
\\
0&{i\sigma_2}
\end{pmatrix}
\equiv\frac{1}{\Lambda_{\rm NC}^2}i\sigma_2\otimes I_2,
\label{nondegen}
\end{gather}
with $\sigma_2$ being famous Pauli matrix.
This constraints was used in the renormalizability studies of 4d NCGFT without SW map~\cite{Blaschke:2010ck, Blaschke:2009aw}.
Note also that this $\theta^{\mu\nu}_{\sigma_2}$ is full rank and thus breaks in general the unitarity if one performs Wick
rotation to the Minkowski space\-time~\cite{Gomis:2000zz}.
This choice, in 4d Euclidean space\-time, induces a~relation $(\theta\theta)^{\mu\nu}=-\frac{1}{\Lambda_{\rm NC}^4}g^{\mu\nu}$.
The tensor structures~\eqref{PlD}, with restored coupling constant~$e$ included then reduces into two parts and we obtain the
same Lorentz structure as we did from the fermion-loop~\eqref{Flfinal}:
\begin{gather*}
\Pi_{\kappa_g}^{\mu\nu}(p)_4\Big|^{\theta_{\sigma_2}}
=\frac{e^2}{(4\pi)^2}\big\{\big[g^{\mu\nu}p^2-p^\mu p^\nu\big]B_{\rm a}^{\kappa_g}(p)
+(\theta p)^\mu(\theta p)^\nu B_{\rm b}^{\kappa_g}(p)\big\}.
\end{gather*}
Neglecting the IR safe terms we have found that the $B_{\rm a}^{\kappa_g}(p)$ and $B_{\rm b}^{\kappa_g}(p)$ exhibits {\it
divergent} structures~\cite{Horvat:2013rga}:
\begin{gather*}
B_{\rm a}^{\kappa_g}(p)\sim\frac{4(\kappa_g-3)^2}{3}\bigg(\frac{2}{\epsilon}+\ln\big(\mu^2(\theta
p)^2\big)\bigg)+\frac{16}{3}\frac{(\kappa_g-3)(\kappa_g+1)}{p^2(\theta p)^2},
\\
B_{\rm b}^{\kappa_g}(p)\sim2p^2\frac{(\kappa_g-3)(7\kappa_g-9)}{(\theta p)^2}\bigg(\frac{2}{\epsilon}+\ln\big(\mu^2(\theta
p)^2\big)\bigg)-\frac{16}{3}\frac{(\kappa_g-3)(7\kappa_g-5)}{(\theta p)^4},
\end{gather*}
which can all be eliminated by choosing the deformation freedom point $\kappa_g=3$.
A careful evaluation of the full photon-loop at this point exhibits a~simple structures
\begin{gather*}
B_{\rm a}^{\kappa_g=3}(p) =2\bigg[\frac{56}{3}+I\bigg],
\qquad
B_{\rm b}^{\kappa_g=3}(p) =-\frac{p^2}{(\theta p)^2}9[8-I],
\end{gather*}
where in~\cite{Horvat:2013rga} we have shown that
\begin{gather*}
I=0.
\end{gather*}
Thus, for special choice~\eqref{nondegen} in the $D\to 4-\epsilon$ limit, and at $\kappa_g=3$ point, we have found
\begin{gather*}
B_{\rm a}^{\kappa_g=3}(p)=\frac{112}{3},
\qquad
B_{\rm b}^{\kappa_g=3}(p)=-72\frac{p^2}{(\theta p)^2}.
\end{gather*}

Summing up the contributions to the photon polarization tensor.
To simplify the tremendous divergent structures in the loop-coef\/f\/icients $F^{\kappa_f}_{1,2}(p)$'s, and
$B^{\kappa_g}_{1,\dots,5}(p)$'s at $D\to 4-\epsilon$, we have been forced to probe two additional constraints: One which appears
to be inef\/fective is the zero mass-shell condition/limit $p^2\to0$, due to the uncertainty on its own validity when quantum
corrections present.
The other constraint, namely setting $\theta^{\mu\nu}$ to a~special full ranked value
$\theta^{\mu\nu}_{\sigma_2}$~\eqref{nondegen}, reduces the number of dif\/ferent momentum structures from f\/ive to two.
Then all divergences and the IR safe contributions disappear at a~deformation parameter-space unique point
$(\kappa_f,\kappa_g)=(0,3)$ leaving,
\begin{gather*}
\Pi^{\mu\nu}_{(0,3)}(p)\big|^{\theta_{\sigma_2}}\equiv\big[\Pi^{\mu\nu}_{\kappa_f=0}(p)+\Pi^{\mu\mu}_{\kappa_g=3}(p)\big]\big|^{\theta_{\sigma_2}}
= \frac{e^2p^2}{\pi^2}\bigg[\frac{7}{3}\bigg(g^{\mu\nu}-\frac{p^\mu p^\nu}{p^2}\bigg) -\frac{9}{2}\frac{(\theta p)^\mu (\theta
p)^\nu}{(\theta p)^2}\bigg],
\end{gather*}
as the only one-loop-f\/inite contribution/correction to the photon two-point function.

\section{Neutrino two-point function}\label{Section4}

\begin{figure}[t]\centering
\includegraphics[width=5.5cm]{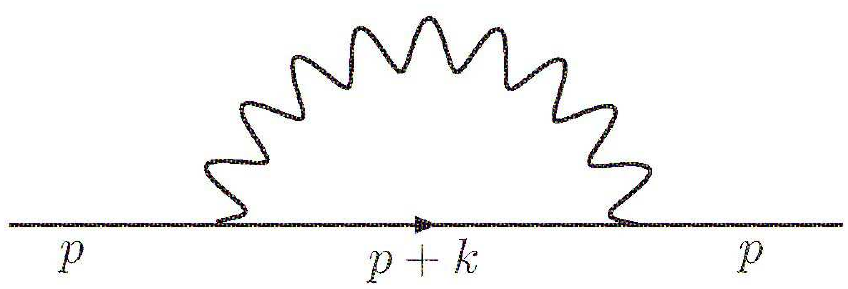}
\caption{Bubble-graph contribution to the neutrino two-point function.}
\label{FigFydiagr3}
\end{figure}

One-loop contributions as a~function of $\kappa_f$ receive the same spinor structure as in~\cite{Horvat:2011qg}.
We now reconf\/irm that by using the action~\eqref{S} together with the Feynman rule~\eqref{Ff}, out of four diagrams in Fig.~2
of~\cite{Horvat:2011qg}, only the non-vanishing bubble-graph is considered in this manuscript.

Thus, in the present scenario a~contribution from Fig.~\ref{FigFydiagr3} reads
\begin{gather}
\Sigma_{\kappa_f}(p)_D=\frac{-1}{(4\pi)^2}\big[\gamma_\mu p^\mu N_1^{\kappa_f}(p)+\gamma_\mu(\theta\theta p)^\mu
N^{\kappa_f}_2(p)\big].
\label{Sigma1}
\end{gather}
Complete loop-coef\/f\/icients $N^{\kappa_f}_{1,2}(p)$, as a~functions of an arbitrary dimensions~$D$ are given
in~\cite{Horvat:2013rga}.

For arbitrary $\kappa_f$ in the limit $D\to 4-\epsilon$ we have obtained the following loop-coef\/f\/icients:
\begin{gather*}
N_1^{\kappa_f}(p)= \kappa_f\bigg\{\bigg[2\kappa_f+\big(\kappa_f-1\big)\bigg(\frac{2}{\epsilon}+\ln{\pi
e^{\gamma_{\rm E}}}+\ln\big(\mu^2(\theta p)^2\big)\bigg)\bigg]
\\
\phantom{N_1^{\kappa_f}(p)=}
{}-\frac{p^2(\theta p)^2}{4}\sum\limits_{k=0}^\infty\frac{(p^2(\theta p)^2)^k}{4^kk(k+1)(2k+1)^2(2k+3)\Gamma[2k+4]}
\\
\phantom{N_1^{\kappa_f}(p)=}
\times
\big[k(k+1)(2k+1)(2k+3)\big(\kappa_f(2k+3)-1\big)\big(\ln\big(p^2(\theta p)^2\big)-2\psi(2k)-\ln4\big)
\\
\phantom{N_1^{\kappa_f}(p)=}
{}+3+28k+46k^2+20k^3-\kappa_f\big(2k+3\big)^2\big(1+8k+8k^2\big)\big]
\\
\phantom{N_1^{\kappa_f}(p)=}
{}+(\tr\theta\theta)\bigg\{\frac{p^2}{(\theta p)^2}\bigg[\frac{2}{\epsilon}+2+\gamma_{\rm E}+\ln\pi+\ln\big(\mu^2(\theta
p)^2\big)+\frac{8\big(\kappa_f-1\big)}{3\kappa_f(\theta p)^2p^2}\bigg]
\\
\phantom{N_1^{\kappa_f}(p)=}
{}-\frac{p^4}{4}\sum\limits_{k=0}^\infty\frac{(p^2(\theta p)^2)^k}{4^kk(k+1)(2k+1)^2(2k+3)}\big[k(k+1)(2k+1)(2k+3)
\\
\phantom{N_1^{\kappa_f}(p)=}
\times
\big(\ln\big(p^2(\theta p)^2\big)-2\psi(2k)-\ln4\big)-2k\big(14+k(23+10k)\big)-3\big]\bigg\}
\\
\phantom{N_1^{\kappa_f}(p)=}
{}+(\theta\theta p)^2\bigg\{2\frac{p^2}{(\theta p)^4}\bigg[\frac{2}{\epsilon}+1+\gamma_{\rm E}+\ln\pi+\ln\big(\mu^2(\theta
p)^2\big)+\frac{16\big(\kappa_f-1\big)}{3\kappa_f(\theta p)^2p^2}\bigg]
\\
\phantom{N_1^{\kappa_f}(p)=}
{}+\frac{p^4}{2(\theta p)^2}\sum\limits_{k=0}^\infty\frac{k(p^2(\theta p)^2)^k}{(k+1)(2k+1)^2(2k+3)\Gamma[2k+4]}\big[(k+1)(2k+1)(2k+3)
\\
\phantom{N_1^{\kappa_f}(p)=}
\times
\big(\ln\big(p^2(\theta p)^2\big)-2\psi(2k)-\ln4\big)+16k^2-34k-17\big]\bigg\}\bigg\},
\\
N_2^{\kappa_f}(p)= -\kappa_f\frac{p^2}{(\theta p)^2}\bigg\{4+\big(\kappa_f-1\big)\bigg[\frac{2}{\epsilon}+\ln{\pi
e^{\gamma_{\rm E}}}+\ln\big(\mu^2(\theta p)^2\big)\bigg]-\frac{16\big(\kappa_f-1\big)}{3(\theta p)^2p^2}
\\
\phantom{N_2^{\kappa_f}(p)=}
{}-\frac{p^2(\theta p)^2}{4}\sum\limits_{k=0}^\infty\frac{(p^2(\theta p)^2)^k}{4^kk(k+1)(2k+1)^2(2k+3)\Gamma[2k+4]}
\big[k(1+k)(1+2k)(3+2k)
\\
\phantom{N_2^{\kappa_f}(p)=}
\times
\big(1+3k_f+2\big(k_f+1\big)k\big)\big(\ln\big(p^2(\theta p)^2\big)-2\psi(2k)-\ln4\big)-3-9\kappa_f
\\
\phantom{N_2^{\kappa_f}(p)=}
{}-4k\big(7+21\kappa_f+\big(20+43\kappa_f +2k\big(11+4k+4\kappa_f(4+k)\big)\big)k\big)\big]\bigg\}.
\end{gather*}
In the expressions for $N_{1,2}^{\kappa_f}(p)$ contributions from both the planar as well as the non-planar graphs are present.
For any $\kappa_f\not=1$ our neutrino two point function receive UV, and power as well as logarithmic UV/IR mixing terms.

{\bf First we analyze the choice $\boldsymbol{\kappa_f=1}$}.
For $D=4-\epsilon$ in the limit $\epsilon\to 0$, we obtain the f\/inal expression as
\begin{gather}
\Sigma_{\kappa_f=1}(p)=\frac{-e^2}{(4\pi)^2} \gamma_{\mu} \bigg[p^{\mu} N_1(p)+(\theta{\theta p})^{\mu}\frac{p^2}{(\theta
p)^2}N_2(p)\bigg],
\label{sigma1AB}
\end{gather}
with restored coupling constant~$e$ included.
Here $N_{1,2}(p)$ coef\/f\/icients are as follows
\begin{gather}
N_1(p)= p^2\bigg(\frac{\tr\theta\theta}{(\theta p)^2} +2\frac{(\theta\theta p)^2}{(\theta p)^4}\bigg) A
+\bigg[1+p^2\bigg(\frac{\tr\theta\theta}{(\theta p)^2} +\frac{(\theta\theta p)^2}{(\theta p)^4}\bigg)\bigg]B,
\\
A=\frac{2}{\epsilon}+\ln(\mu^2(\theta p)^2) + \ln({\pi e^{\gamma_{\rm E}}})
+\sum\limits_{k=1}^\infty \frac{\left(p^2(\theta p)^2/4\right)^k}{\Gamma(2k+2)} \left(\ln\frac{p^2(\theta p)^2}{4}+2\psi_0(2k+2)\right),
\\
B= -8\pi^2 N_2(p) = -2
+\sum\limits_{k=0}^\infty \frac{\left(p^2(\theta p)^2/4\right)^{k+1}}{(2k+1)(2k+3)\Gamma(2k+2)}
\nonumber
\\
\phantom{B=}
{}\times\left(\ln\frac{p^2(\theta p)^2}{4}- 2\psi_0(2k+2) - \frac{8(k+1)}{(2k+1)(2k+3)} \right).
\label{A2}
\end{gather}
It is to be noted here that the spinor structure proportional to ${\gamma_{\mu}(\theta p)^{\mu}}$ is missing in the f\/inal
result.
This conforms with the calculation of the neutral fermion self-energy in the $\theta$-expanded SW map
approach~\cite{Ettefaghi:2007zz}.

The $1/\epsilon$ UV divergence could in principle be removed by a~properly chosen counterterm.
However (as already mentioned) due to the specif\/ic momentum-dependent coef\/f\/icient in front of it, a~nonlocal form for it is
required\footnote{Any quadratic counterterm for the neutral fermion would be gauge invariant since the neutral fermion f\/ield is
invariant under the U(1) gauge transformation.}.
It is important to stress here that amongst other terms contained in both coef\/f\/icients $N_{1,2}(p)$, there are structures
proportional to
\begin{gather*}
\big({p^2(\theta p)^2}\big)^{n+1}\big(\ln\big(p^2(\theta p)^2\big)\big)^m,
\qquad
{\forall\, n}
\quad
{\rm and}
\quad
m=0,1.
\end{gather*}
The numerical factors in front of the above structures are rapidly-decaying, thus series are always convergent for f\/inite
argument, as we demonstrate in~\cite{Horvat:2011qg}.

Turning to the UV/IR mixing problem, we recognize a~soft UV/IR mixing term represented by a~logarithm,
\begin{gather*}
\Sigma_{\kappa_f=1}^{\rm UV/IR}=\frac{-e^2}{(4\pi)^2}{\fmslash p}p^2  \bigg(\frac{\tr\theta\theta}{(\theta p)^2}
+2\frac{(\theta\theta p)^2}{(\theta p)^4}\bigg)\cdot \ln{\big|\mu^2(\theta p)^2\big|}.
\end{gather*}
Instead of dealing with nonlocal counterterms, we take a~dif\/ferent route here to cope with various divergences
besetting~\eqref{sigma1AB}.
Since $\theta^{0i} \neq 0$ makes a~NC theory nonunitary~\cite{Gomis:2000zz}, we can, without loss of generality, chose $\theta$
to lie in the (1, 2) plane
\begin{gather}
\theta^{\mu\nu}_{\rm spec}=\frac{1}{\Lambda_{\rm NC}^2}
\begin{pmatrix}
0&0&0&0
\\
0&0&1&0
\\
0&-1&0&0
\\
0&0&0&0
\end{pmatrix}
,
\label{degen}
\end{gather}
yielding
\begin{gather}
\frac{\tr\theta\theta}{(\theta p)^2} +2\frac{(\theta\theta p)^2}{(\theta p)^4}=0,
\quad
\forall\, p.
\label{degen1}
\end{gather}
With~\eqref{degen1}, $\Sigma^{\rm spec}_{\kappa_f=1}$, in terms of Euclidean momenta, receives the following form:
\begin{gather}
\Sigma^{\rm spec}_{\kappa_f=1}(p)=\frac{-e^2}{(4\pi)^2}\gamma_\mu\left[p^\mu \bigg(1 + \frac{\tr\theta\theta}{2}\frac{p^2}{(\theta
p)^2}\bigg) -2(\theta\theta p)^\mu\frac{p^2}{(\theta p)^2} \right] N_2(p).
\label{Sigmabuble}
\end{gather}
By inspecting~\eqref{A2} one can be easily convinced that $N_2(p)$ is free from the $1/\epsilon$ divergence and the UV/IR mixing
term, being also well-behaved in the infrared, in the $\theta \rightarrow 0$ as well as $\theta p \rightarrow 0$ limit.
We see, however, that the two terms in~\eqref{Sigmabuble}, one being proportional to $p^\mu$ and the other proportional to
$(\theta\theta p)^\mu$, are still ill-behaved in the $\theta p \rightarrow 0$ limit.
If, for the choice~\eqref{degen},~$P$ denotes the momentum in the (1,2) plane, then $\theta p = \theta P$.
For instance, a~particle moving inside the noncommutative plane with momentum~$P$ along the one axis, has a~spatial extension of
size~$|\theta P|$ along the other.
For the choice~\eqref{degen}, $\theta p \rightarrow 0$ corresponds to a~zero momentum projection onto the (1,2) plane.
Thus, albeit in our approach the commutative limit~($\theta \rightarrow 0$) is smooth at the quantum level, the limit when an
extended object (arising due to the fuzziness of space) shrinks to zero, is not.
We could surely claim that in our approach the UV/IR mixing problem is considerably softened; on the other hand, we have
witnessed how the problem strikes back in an unexpected way.
This is, at the same time, the f\/irst example where this two limits are not degenerate (or two limits do not commute).

{\bf Next we analyze the choice $\boldsymbol{\kappa_f=0}$}.
Using the Feynman rule~\eqref{Ff} for $\kappa_f=0$ and for general $\theta$, we f\/ind the following closed form contribution to
the neutrino two point function (from diagram $\Sigma_1$ in~\cite{Horvat:2011qg, Horvat:2011bs}):
\begin{gather}
\Sigma_{\kappa_f=0}(p)=\frac{e^2}{(4\pi)^2} {\fmslash p}\bigg[\frac{8}{3} \frac{1}{{(\theta p)^2}}
\bigg(\frac{\tr\theta\theta}{(\theta p)^2} +4\frac{(\theta\theta p)^2}{(\theta p)^4}\bigg)\bigg] .
\label{A9}
\end{gather}
It is important to stress that we have found that diagram~$\Sigma_2=0$ in all $\kappa_f$-cases, while diag\-rams~$\Sigma_3$ and~$\Sigma_4$ vanish due to charge conjugation symmetry, see~\cite{Horvat:2011qg, Horvat:2011bs}.
There is no alternative dispersion relation in degenerate case~\eqref{degen}, since the factor that multiplies $\fmslash p$
in~\eqref{A9}, does not dependent on the time-like component $p_0$ (energy).

Considering neutrino two point function~\eqref{Sigma1}, our results extends the prior works~\cite{Horvat:2011qg, Horvat:2011bs}
by completing the behavior for general~$\kappa_f$.
Here we discuss some novel behaviors associated with general~$\kappa_f$.
The neutrino two point function does posses power UV/IR mixing phenomenon for arbitrary values of~$\kappa_f$, except
$\kappa_f=1$.
In the limit $\kappa_f\to 0$ all UV, IR divergent terms as well as constant terms in $N^{\kappa_f}_{1,2}(p)$ vanish; what
remains are only the power UV/IR mixing terms.
The UV divergence can be localized using the special~$\theta$ value~\cite{Horvat:2011qg, Horvat:2011bs} in $N^{\kappa_f}_1(p)$
but not in $N^{\kappa_f}_2(p)$.
The UV and the power IR divergence in $N^{\kappa_f}_2(p)$ can be removed by setting $\kappa_f=1$.

Summing up, choice $\kappa_f=1$ eliminates some of divergences, but not all of them.
Imposing the special $\theta^{\mu\nu}_{\sigma_2}$ reduces the contribution to quadratic UV/IR mixing into a~single term from
$N_2^{\kappa_f}(p)$, which has two zero points $\kappa_f=0,1$.
Only $\kappa_f=0$ can induce full divergence cancelations, by removing the whole $\Sigma(p)$, i.e.\
we have
\begin{gather*}
\Sigma_{\kappa_f=0}(p)\big|^{\theta_{\sigma_2}}=0.
\end{gather*}

\section{Discussion and conclusion}\label{Section5}

In this review we present a~$\theta$-exact quantum one-loop contributions to the photon ($\Pi$) and neutrino ($\Sigma$) two
point functions and analyze their properties.
In principle the quantum corrections in NCQFTs are extremely profound, revealing a~structure of pathological terms far beyond
that found in ordinary f\/ield theories.
For practical purposes, perturbative loop computation was the most intensively studied for the Moyal--Weyl (constant $\theta^{\mu
\nu}$) type deformation~\cite{Blaschke:2010ck, Blaschke:2009aw,Filk:1996dm,Grosse:2004yu,Magnen:2008pd}, for its preservation of
translation invariance allows a~(modif\/ied) Feynman diagrammatic calculation.
Much of ef\/forts went in taming divergences related to the ultraviolet-infrared connection/mixing, see for
example~\cite{Bigatti:1999iz,Horvat:2011qg,Horvat:2011bs,Horvat:2013rga,Matusis:2000jf,Minwalla:1999px}.
The UV/IR mixing, built-in as a~new principle in all NCQFT models, and closely related to the Black Hole Complementarity and/or
Holographic principle~\cite{Horvat:2010km,Susskind:2005js,Susskind:1993if,'tHooft:1984re}, does reverse the well-established
connection (via the uncertainty principle) between energy and size.
The existence of such pathological terms in NCQFTs raises a~serious concern on the renormalizability/consistency of the theory.
Although no satisfactory resolution for this issue had been achieved so far, very recently it has been observed that certain
control over novel divergences may be obtained by certain gauge invariant variation of the SW mapped
action~\cite{Horvat:2013rga}.
The anomalous structures in the two point function further suggests possible modif\/ications to obtain trouble-free and meaningful
loop-results, necessary for stu\-dying the particle propagation~\cite{Horvat:2011qg}.
Such ef\/fects were largely left untouched in literature so far, mostly due to the prior concern on the
consistency/renormalizability from a~purely theoretical viewpoint.

After having def\/ined and explained the full noncommutative action-model origin of the deformation parameter-space $(\kappa_f,
\kappa_g)$, we obtained the relevant Feynman rules.
Our method, extending the modif\/ied Feynman rule procedure~\cite{Filk:1996dm}, yields the one-loop quantum corrections for
arbitrary dimensions in closed form, as function of the deformation-freedom parameters~$\kappa_f$,~$\kappa_g$, as well as momentum
$p^\mu$ and noncommutative parameter~$\theta^{\mu\nu}$.
Full parameter-space freedom is kept in our evaluation here.
Following the extended dimensional regularization technique we expressed the diagrams as~$D$-dimensional loop-integrals and
identify the relevant momentum structures with corresponding loop-coef\/f\/icients.
We have found that total contribution to photon two-point function satisf\/ies the Ward--(Slavnov--Taylor) identity for arbitrary
dimensions~$D$ and for any point in the $(\kappa_f,\kappa_g$) parameter-space:
\begin{gather*}
p_\mu\Pi_{(\kappa_f,\kappa_g)}^{\mu\nu}(p)_{D}=p_\mu\big(\Pi_{\kappa_f}^{\mu\nu}(p)_{D}+\Pi_{\kappa_g}^{\mu\nu}(p)_{D}\big)
=p_\nu\big(\Pi_{\kappa_f}^{\mu\nu}(p)_{D}+\Pi_{\kappa_g}^{\mu\nu}({p})_{D}\big)=0.
\end{gather*}

The one-loop photon polarization tensor in four dimensions contains the UV divergence and UV/IR mixing terms dependent on the
freedom parameters $\kappa_f$ and $\kappa_g$.
The introduction of the freedom parameters univocally has a~potential to improve the situation regarding cancellation of
divergences, since certain choices for $\kappa_f$ and $\kappa_g$ could make some of the terms containing singularities to vanish.

We observe the following general behavior of one-loop two-point functions in the $D\to 4-\epsilon$ limit: The total expressions
for both the photon and the neutrino two point functions contain the $1/\epsilon$ ultraviolet term, the celebrated UV/IR mixing
power terms as well as the logarithmic (soft) UV/IR mixing term.
The $1/\epsilon$ divergence is always independent of the noncommutative scale.
The logarithmic terms from the $\epsilon$-expansion and the modif\/ied Bessel function integral sum into a~common term
$\ln(\mu^2(\theta p)^2)$, which is divergent in the infrared $|p|\to 0$, in the size-of-the-object $|\theta p|\to 0$ limit, as
well as in the vanishing noncommutativity $\theta \to 0$ limit.
Thus, the existences of UV/IR mixings for both, photons and neutrinos respectively, in 4d spaces deformed by spacetime
noncommutativity at low energies, suggests that the relation of quantum corrections to
observations~\cite{hep-th/0606248,Abel:2006wj,AlvarezGaume:2003mb, Horvat:2010km,Jaeckel:2005wt} is not entirely clear.
However, in the context of the UV/IR mixing it is very important to mention a~complementary
approach~\cite{hep-th/0606248,Abel:2006wj} where NC gauge theories are realized as ef\/fective QFT's, underlain by some more
fundamental theory such as string theory.
It was claimed that for a~large class of more general QFT's above the UV cutof\/f the phenomenological ef\/fects of the UV
completion can be quite successfully modeled by a~threshold value of the UV cutof\/f.
So, in the presence of a~f\/inite UV cutof\/f no one sort of divergence will ever appear since the problematic phase factors
ef\/fectively transform the highest energy scale (the UV cutof\/f) into the lowest one (the IR cutof\/f).
What is more, not only the full scope of noncommutativity is experienced only in the range delimited by the two cutof\/fs, but for
the scale of NC high enough, the whole standard model can be placed below the IR cutof\/f~\cite{Horvat:2010km}.
Thus, a~way the UV/IR mixing problem becomes hugely less pressing, making a~study of the theory at the quantum level much more
reliable.

We have demonstrated how quantum ef\/fects in the $\theta$-exact Seiberg--Witten map approach to NC gauge f\/ield theory reveal
a~much richer structure for the one-loop quantum correction to the photon and fermion two-point functions (and accordingly for
the UV/IR mixing problem) than observed previously in approximate models restricting to low-energy phenomena.
Our analysis can be considered trustworthy since we have obtained the f\/inal result in an analytic, closed-form manner.
We believe that a~promising avenue of research would be using the enormous freedom in the Seiberg--Witten map to look for other
forms which UV/IR mixing may assume.
Two alternative forms have been already found~\cite{Horvat:2011qg}.
Finally, our approach to UV/IR mixing should not be confused with the one based on a~theory with UV completion ($\Lambda_{\rm
UV} < \infty$), where a~theory becomes an ef\/fective QFT, and the UV/IR mixing manifests itself via a~specif\/ic relationships
between the UV and the IR cutof\/fs~\cite{hep-th/0606248,Abel:2006wj,AlvarezGaume:2003mb, Horvat:2010km,Jaeckel:2005wt}.

In conclusion, our main result in four-dimensional space is that we have all pathological terms under full control after the
introduction of the deformation-freedom parameter-space $(\kappa_f,\kappa_g)$ and a~special choice for $\theta^{\mu\nu}$.
Namely, working in the 4d Euclidean space with a~special full rank of $\theta^{\mu\nu}_{\sigma_2}$ and setting
the point $(\kappa_f,\kappa_g)=(0,3)$, the fermion plus the photon-loop contribution to $\Pi^{\mu\nu}_{(\kappa_f,\kappa_g)}(p)$ contain
only two f\/inite terms, i.e.\
all divergent terms are eliminated.
In this case the neutrino two-point function vanishes.

\subsection*{Acknowledgments}

J.T.~would like to acknowledge J.~Erdmenger and W.~Hollik for discussions, and Max-Planck-Institute for Physics, Munich, for
hospitality.
A great deal of computation was done by using {\sc Mathematica}~8.0\footnote{Wolfram Research Inc., Mathematica, Version 8.0, Champaign, IL, 2010.} and tensor algebra package
xAct\footnote{Martin-Garcia J., xAct, \url{http://www.xact.es/}}.

\pdfbookmark[1]{References}{ref}
\LastPageEnding

\end{document}